\documentclass[conference]{IEEEtran}
\ifCLASSINFOpdf
  \usepackage[pdftex]{graphicx}
\else
  \usepackage[dvips]{graphicx}
\fi
\usepackage[cmex10]{amsmath}
\usepackage{url}
\usepackage{algorithm}
\usepackage{algorithmic}
\usepackage{subfigure}

\IEEEoverridecommandlockouts
\begin{document}
%\title{Design of Structured Random Linear Codes for AL-FEC Block and Convolutional Codes} 
\title{Structured Random Linear Codes (SRLC): Bridging the Gap between Block and Convolutional Codes} 

%------------------------------------------------------------------
\author{\IEEEauthorblockN{Kazuhisa Matsuzono\IEEEauthorrefmark{1},
Vincent Roca\IEEEauthorrefmark{2} and
Hitoshi Asaeda\IEEEauthorrefmark{1}}
\IEEEauthorblockA{\IEEEauthorrefmark{1}National Institute of Information and Communications Technology (NICT),\\
4-2-1, Nukui-Kitamachi Koganei, Tokyo 184-8795, Japan\\ Email: \{matsuzono, asaeda\}@nict.go.jp
\IEEEauthorblockA{\IEEEauthorrefmark{2}Inria, Privatics team, Grenoble, France\\
Email: vincent.roca@inria.fr}}
\thanks{This work was supported in part by the ANR-09-VERS-019-02 grant (ARSSO project).
This work has been carried out in part while K. Matsuzono was visiting Inria (France) as a Post-Doctoral fellow.}
}

\maketitle

\begin{abstract}
Several types of AL-FEC (Application-Level FEC) codes for the Packet Erasure Channel exist.
Random Linear Codes (RLC), where redundancy packets consist of random linear combinations of source packets over a certain finite field, are a simple yet efficient coding
technique, for instance massively used for Network Coding applications.
However the price to pay is a high encoding and decoding complexity, especially when working on $GF(2^8)$, which seriously limits the number of packets in the encoding window.
On the opposite, structured block codes have been designed for situations where the set of source packets is known in advance, for instance with file transfer applications.
Here the encoding and decoding complexity is controlled, even for huge block sizes, thanks to the sparse nature of the code and advanced decoding techniques that exploit
this sparseness (e.g., Structured Gaussian Elimination).
But their design also prevents their use in convolutional use-cases featuring an encoding window that slides over a continuous set of incoming packets.

In this work we try to bridge the gap between these two code classes, bringing some structure to RLC codes in order to enlarge the use-cases where they
can be efficiently used: in convolutional mode (as any RLC code), but also in block mode with either tiny, medium or large block sizes.
We also demonstrate how to design compact signaling for these codes (for encoder/decoder synchronization), which is an essential practical aspect.
\end{abstract}

\IEEEpeerreviewmaketitle

%==================================================================
\section{Introduction}
\label{sec:intro}
%%%% describe AL-FEC overview and the reason why it's important.
Application-Level Forward Erasure Correction (AL-FEC) codes have become a key component of many content delivery systems.
They are widely used as an efficient technique to recover packet losses in Internet (usually caused by congested routers) or wireless communications (often caused by a short term fading problem).
Such a network can be regarded as a packet erasure channel (or equivalently a Bit Erasure Channel, BEC), characterized by the property that the transmitted data packets are either received without error or erased (lost).
Packet loss resilience may also be achieved with Automatic Repeat reQuest (ARQ) techniques (e.g., with TCP), but:
a Round Trip Time (RTT) is needed to recover from a loss, which can be a issue for delay-sensitive applications (e.g., video-conferencing),
the return channel may not exist (e.g., in case of a unidirectional broadcast network), and 
it does not scale well with the number of receivers in case of multicast or broadcast transmissions.

AL-FEC codes are a key building block of content broadcast technologies such as the FLUTE/ALC~\cite{flute} protocol stack 
for the reliable and scalable transmission of files to a potentially huge number of receivers, and the FECFRAME framework~\cite{fecframework} when dealing with
real-time delivery services as in streaming applications.
AL-FEC is now deployed in all systems relying on FLUTE/ALC (e.g., 3GPP MBMS service~\cite{3GPP-MBMS} or ISDB-Tmm~\cite{ISDB-Tmm}) and sometimes at the link layer as well
(e.g., the MPE-FEC layer of DVB-H systems~\cite{MPE-FEC}).

In all of the previous use-cases (real-time delivery included), only block codes are considered, and the set of source packets is first grouped into blocks where AL-FEC encoding/decoding is performed.
In the present work we introduce an alternative and practical AL-FEC solution that aims at encompassing both block oriented and sliding window oriented use-cases.

%%%%% describe 1) RLC Overview. 2) explain why we focus on RLC ? and it's needed and important ?
Random Linear Codes (RLC) are another class of AL-FEC codes.
They are increasingly popular due to their simple yet powerful encoding techniques, in particular in the context of Random Linear Network Coding (RLNC) where encoding/decoding can be performed at the various network nodes, namely either at intermediate nodes (e.g., WiFi Access point or routers) or end nodes~\cite{rlnc}.
At a source (sender) node or intermediate node, RLC generates encoded packets (also called encoded symbols in this paper) just by linearly combining the available symbols using encoding vectors (also called coefficients) randomly selected from a given finite field (e.g., $GF(2^8)$).
In general, the set of available symbols evolves over the time, i.e., RLC are used as convolutional codes.
%RLC can also be utilized for convolutional coding where the past encoded source symbols have the possibility to be encoded multiple times.
%As in~\cite{tetrys}, RLC is utilized in a convolutional mode, in an end-to-end manner (without any re-encoding within the core network), with feedback information from a receiver, which makes it possible to achieve a full (or partial) reliability, which is RTT-independent.
In~\cite{tetrys}, RLC is also utilized in a convolutional manner, but end-to-end (i.e., there is no re-encoding within the core network), with feedback information from the receiver, which enables to achieve a full reliability when desired.
The authors show that the recovery delay for lost packets is in that case independent of the RTT.
However the main issue to be considered with RLC is the high decoding complexity, typically a Gaussian Elimination (GE) over a dense linear system.
This problem becomes even more pronounced when the number of source symbols involved is large, and/or when the finite field is $GF(2^8)$ (or higher) so as to improve the erasure recovery capabilities~\cite{rlc-raptor-comp}.

%%%% describe 1) our motivations; there are incentives to have ``structured'' codes. 2) our goals in this work
Since we believe that RLC can play a key role in network coding systems for the erasure channel~\cite{nwcrg}, we have focused on the design of new improved RLC techniques.
Our goal is to design RLC codes that:
\begin{itemize}
\item can be used either as block or convolutional codes;
\item can be used with encoding window sizes in 2--10,000s symbols range, as very large sizes are beneficial to bulk file transfers while small values are useful for real-time contents;
\item have excellent erasure recovery performance, and at the same time enable fast encoding/decoding which is essential for devices with limited computational and memory capabilities;
\item enable compact signaling (e.g., transmitting the full encoding vector does not scale); % with the encoding window size);
%\item and does not use patented techniques so as to become an IPR free solution as much as possible.	% VR: it is risky to claim we are IPR-free in a research paper...
\end{itemize}
In other words, we try to bridge the gap between block and convolutional AL-FEC codes.
With these goals in mind, we have designed the so-called Structured Random Linear Codes (SRLC) \cite{nsrlc.ietf88}.% that take advantage of a specific structure that is added to otherwise largely sparse codes (just like the Repeat-Accumulate structure added to sparse LDPC codes in LDPC-Staircase~\cite{ldpc-rfc}).
In the present work, as a first step to a complete evaluation, we only focus on use-cases that require only end-to-end encoding (i.e., there is a single end point for AL-FEC encoding/decoding, no matter whether this end is a ``host'' or a ``middlebox'') and we evaluate the SRLC effectiveness in terms of erasure recovery performance only.

%%%% describe the construction of this paper. 
The remainder of this paper is organized as follows: Section~\ref{sec:related-work} introduces related work. Section~\ref{sec:prop-srlc} describes the proposed Structured Random Linear Codes (SRLC) in detail. Section~\ref{sec:eval} evaluates the recovery performances of SRLC, and we conclude in Section~\ref{sec:concl}.

%==================================================================
\section{Related Work}
\label{sec:related-work}
%%%% BAT-codes
In~\cite{BATS1, BATS2} BATched Sparse (BATS) codes are proposed for file distribution through a communication network where intermediate nodes have coding
capabilities~\cite{linear-net-coding}.
These codes are designed so as to control the computational and storage requirements at the source, intermediate nodes, and destination, as well as the transmission overhead when transmitting the coding vector.
This is made possible by the use of both an outer code (sender), that forms "batches" of coded packets, using a specific distribution for defining the number of input packets considered to create each batch, and inner codes (intermediate nodes) that perform random linear codings of packets of a given batch.
The authors show the good performance of these BATCH codes, when associated to "inactivation decoding" as in \cite{vincent-wimob13,raptor}.
%The main goal is to solve high computational and/or storage capacities needed for intermediate nodes.
%BATS codes extend a rateless code approach~\cite{digital-fountain} as an outer code in order to incorporate RLC as inner code. 
%The inner code applies a linear transformation to each ``batched'' encoded packet generated by the outer code.
%The inner code preserves the degrees of the batches to enable an efficient Belief Propagation (BP) decoding and low decoding complexity independent of the total number of packets for transmission (source block length).
%The required computational and storage capacities for the intermediate nodes in a network are independent of the number of packets for transmission.
%As in~\cite{vincent-wimob13, raptor} a Maximum Likelihood (ML) decoding is used with BP decoding so as to fill the gap of the BP decoding failure probability (i.e., BP decoding is not an optimal method for maximum performance). 

%%%% Gamma codes
In~\cite{gamma1, gamma2}, Gamma Codes are proposed as a family of sparse random linear network codes with outer code like BATS codes.
The codes also manage ``chunked'' encoded packets.
The key idea is to enable the outer code to play as soon as the first chunked packets are recovered, which realizes a joint decoder scheme that coordinates a proper combination of an outer coding and a basic sparse random linear network coding.
It was presented that Gamma codes can achieve better reception overhead while keeping lower encoding/decoding complexity in fixed block length configuration.

%%%% differences for related works and future works.
These works differ from ours.
We designed RLC in order to be flexibly used as either block or convolutional codes, over wide ranges of block/encoding-window sizes.
Additionally, we do not distinguish outer/inner codes per se, but add a structure to the RLC approach in order to find an appropriate balance between computational complexity and erasure recovery performance.

%%%% Memo %%%%%
% Concerning Performance evaluations: perform-eval raptor and RLC
% Space codes: Raptor --> BATS
% Mix bin and non-bin: RaptorQ, GLDPC.
% RS+LDPCの論文は、raptorQ, inactive decoding, sparse natureとかの説明につかえるかも。

%==================================================================
\section{Structured Random Linear Codes (SRLC)}
\label{sec:prop-srlc}
Let us now describe the SRLC codes, characterized by:
\begin{enumerate}
\item a mostly sparse binary structure, which reduces the number of symbol XOR operations and improves ITerative (IT)/Structured Gaussian Elimination (SGE) decodings~\cite{itdecoding}\cite{SGE1}\cite{SGE2}. It is a key feature to favor high speed encoding and decoding;
\item a limited use of non-binary (over $GF(2^8)$) coefficients in encoding vectors. It is a key feature to favor a good erasure recovery performance;
\item the addition of a dedicated repeat-and-accumulate structure (as in Irregular Repeat-Accumulate (IRA)~\cite{IRA} and LDPC-Staircase codes). It is a second key feature to favor a good erasure recovery performance.
\end{enumerate}
In the following, we first explain the SRLC approach when used in block mode, and later we extend it to the case of convolutional coding.

%-------------------------------------------------------------------
\subsection{First Idea: Mixing Binary and Non-Binary Coefficients}
\label{subsec:prob-definition}
The first idea consists in using both binary and non-binary coefficients.
All the examples of this section are for the block mode case, when considering a fixed set of $k$ source symbols.

\begin{figure}[h]
    \begin{center}
    \includegraphics[width=220pt]{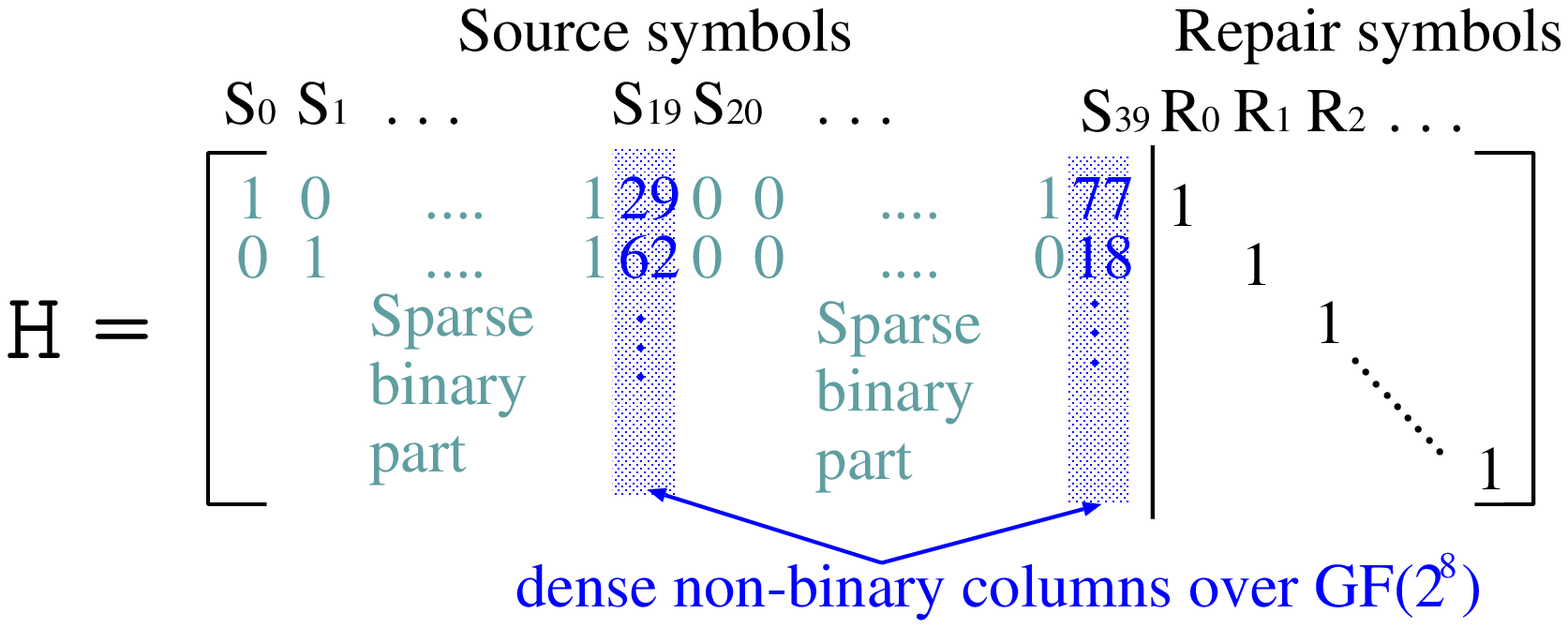}
    \end{center}
    \caption{Example when used as a block code, considering only binary and non-binary coefficients.}
    \label{fig:mix-bin-nonbin}
\end{figure}

%%%% matrixとrepair symbol作成の説明
%Fig.~\ref{fig:mix-bin-nonbin} shows an example of parity check matrix as a block code where the number of source symbols is fixed (40 in the example) that manipulates both binary and non-binary coefficients.
Fig.~\ref{fig:mix-bin-nonbin} shows an example parity check matrix, $H$, when the number of source symbols is fixed (here $k=40$ source symbols) and both binary and non-binary coefficients are used.
The $H$ matrix is composed of two parts, the left side $H_{left}$ and the right side $H_{right}$.
The columns of $H_{left}$ correspond to the source symbols from $S_0$ to $S_{39}$ (A.K.A. source packets), while those of $H_{right}$ correspond to the repair symbols to be generated (A.K.A. repair packets).
Each row of $H_{left}$ represents a constraint (or equation) used for instance to generate the repair symbol of the same row.
For example, the $R_0$ repair symbol in the first row is generated by\footnote{NB: ``$1$'' coefficients are omitted.}:
	$$R_{0}=S_0 + \cdots S_{18}+29*S_{19}+ \cdots S_{38}+77*S_{39}$$

Three key parameters exist:
\begin{itemize}
\item $k$: the source block length (or encoding window size);
\item $D_{bin}$: the density of each ``sparse binary sub-matrix'', given as the ratio of the number of non-zero coefficients to the total number of coefficients in a sub-matrix:
	$$D_{bin} = \frac{\textit{nb\_1\_coeffs}}{\textit{total\_nb\_coeffs\_in\_binary\_submatrix}}$$
\item $D_{nonbin}$: the ratio of the number of non-binary columns to $k$ (i.e., the total number of columns) in $H_{left}$:
	$$D_{nonbin} = \frac{\textit{nb\_nonbinary\_columns}}{k}$$
\end{itemize}

%%%% Sparse parts
$H_{left}$ should be largely composed of sparse binary parts so that most equations are sparse with binary coefficients, because this is a key for high encoding/decoding speeds.
However, a trade-off between speed and erasure recovery performance must be considered since being too sparse and binary negatively impacts the erasure recovery performance.

\begin{figure}[t]
    \begin{center}
    \includegraphics[width=200pt]{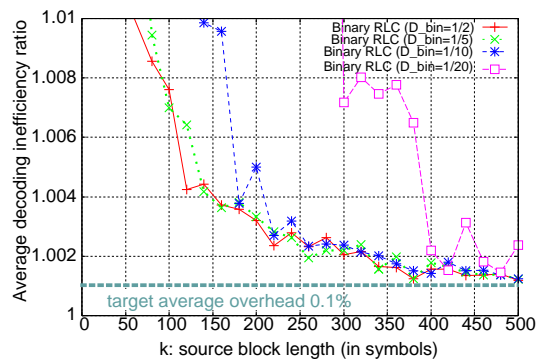}
    \end{center}
	\vspace{-3mm} %XXX
    \caption{Average inefficiency ratio of {\bf binary RLC}, for various binary densities (1/2, 1/5, 1/10, 1/20), as a function of k.}
    \label{fig:avg-ineff-binary-rlc}
\end{figure}

Fig.~\ref{fig:avg-ineff-binary-rlc} shows the average erasure recovery performance of a fully binary RLC with various $D_{bin}$ values as a function of $k$.
The performance metric is the ``average decoding inefficiency ratio'', defined as the ratio of the average number of symbols needed for decoding to complete successfully to $k$:
$$\mathit{inefficiency\_ratio} = (\mathit{nb_{symbols\_needed}}) / k = 1 + \epsilon $$
where $\epsilon$ is called ``decoding overhead'', and also often expressed as a percentage.
The closer to $1$ (achieved with ideal codes) the ratio, the better.
Assuming that our target average decoding overhead is set (arbitrarily) to 0.1\%, we can see in Fig.~\ref{fig:avg-ineff-binary-rlc} that none of the codes achieves the goal, even with binary RLC (where $D_{bin}=1/2$) which performs the best.

\begin{figure}[t]
    \begin{center}
    \includegraphics[width=200pt]{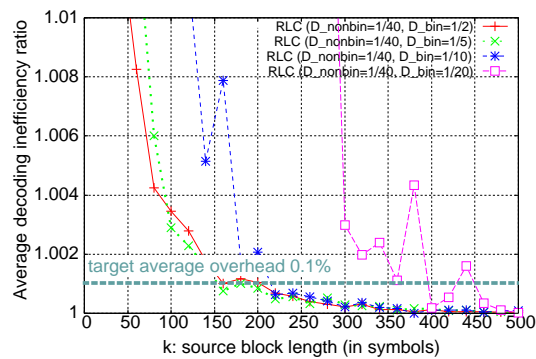}
    \end{center}
	\vspace{-3mm} %XXX
    \caption{Average inefficiency ratio of {\bf RLC with non-binary column (1/40)} with the same binary densities, as a function of k.}
    \label{fig:avg-ineff-nonbin_1_40_-rlc}
\end{figure}

On the opposite, we see in Fig.~\ref{fig:avg-ineff-nonbin_1_40_-rlc} that adding a few dense non-binary columns (we used $D_{nonbin}=1/40$ and the same values for $D_{bin}$), the target average decoding overhead is easily achieved with $k>200$.
And adding more non-binary columns easily enables to further improve the decoding performance with smaller $k$ values.
We will address the question of what are the appropriate \{$D_{nonbin}$, $D_{bin}$\} tuples as a function of $k$ in section~\ref{subsec:add-structure}.

%%%% dense non-binary ``columns''
In the SRLC design, dense non-binary coefficients are always gathered in columns (i.e., assigned to certain symbols).
The motivation is to enable the use of the high speed Structured Gaussian Elimination optimization~\cite{SGE1, SGE2, vincent-wimob13}.
In this approach, when a stopping set is encountered during IT decoding, certain well chosen symbols of the system (i.e., corresponding to unknown/non-recovered source symbols) are logically removed from the linear system.
This process enables IT decoding to pursue.
Finally, decoding finishes with a classic Gaussian Elimination over the removed symbols, and their values are finally re-injected into equations where they were involved.
Because non-binary coefficients are affected to well identified source symbols, these symbols are immediately logically removed from the linear system.
The linear system therefore remains a sparse binary system, not ``polluted'' by non binary coefficients, and most operations consist of fast XOR operations over symbols, a key for high speed decoding.

%-------------------------------------------------------------------
\subsection{Second Idea: Adding a Structure}
\label{subsec:add-structure}
Adding a structure to codes can be highly beneficial.
For instance, the repeat-and-accumulate structure of IRA and LDPC-staircase codes significantly improves their performance: 
because the number of source symbols a repair symbol actually depends on increases with its index, the erasure recovery performance is improved while keeping a sparse system.
However, adding this particular structure would make signaling prohibitively complex when the codes are used in convolutional mode\footnote{
			This is because the encoding window may move in a non predictable way, and in presence of erasures, identifying the exact way all the previously erased
			symbols have been encoded is not easy.}.

\begin{figure}[t]
    \begin{center}
    \includegraphics[width=220pt]{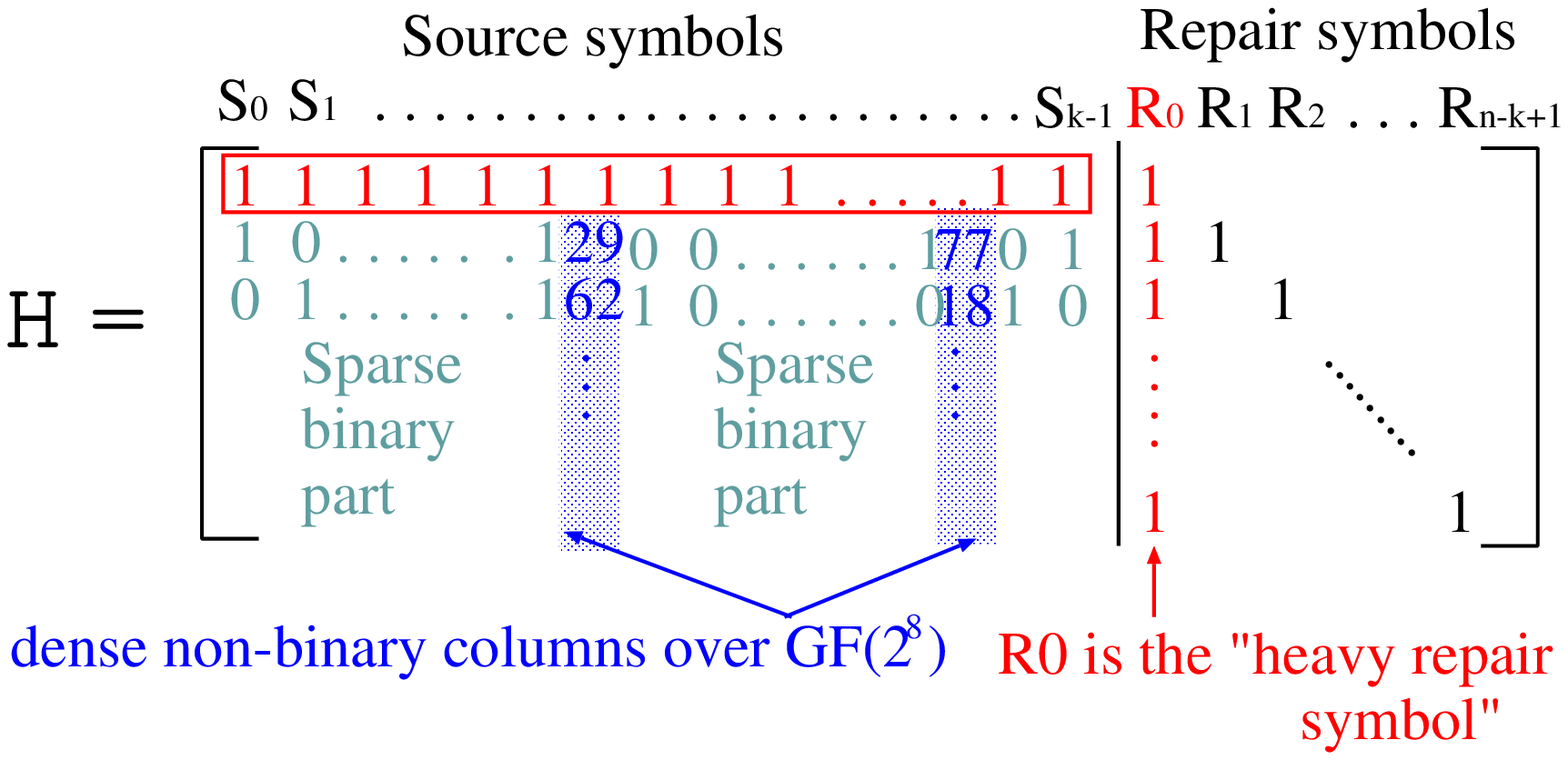}
    \end{center}
	\vspace{-4mm} %XXX
    \caption{SRLC example as a block code.}
    \label{fig:srlc-block-example}
\end{figure}

In order to solve this problem, we propose to add a single accumulative row to create $R_0$, defined as the ``heavy repair symbol'', and to make all repair symbols depend upon $R_0$.
Fig.~\ref{fig:srlc-block-example} shows an example of the proposed SRLC in block mode.
In this example, $R_1$ is generated by:
	$$R_{1}=R_0 + S_0 +...+77*S_{k-3}+S_{k-1}$$

%This simple structure enables a compact signaling thanks to which a receiver determines first the parity check matrix and therefore the relationships between source and repair symbols.
This simple structure enables a compact signaling to enable a receiver to determine the relationships between source and repair symbols. 
This is accomplished, in block mode, by the knowledge of the matrix generation algorithms (specified in a non ambiguous way in the specifications), plus the \{$k$, $D_{nonbin}$, $D_{bin}$\} tuple that is sent once at encoder/decoder synchronization time\footnote{
			This can take several forms, usually in the ``file'' description part of a File Delivery Table (FDT) with FLUTE~\cite{flute}.}.
Because the full encoding vector is not sent along with repair packets (it is useless if $H$ is known), the approach can scale with very large $k$ values.
We will see in section~\ref{subsec:appli-convol} how to perform signaling when SRLC are used in a sliding window mode.

%For instance, as in~\cite{ldpc-rfc}, the use of a function of a key lists (e.g. symbol identifier plus Pseudo-Random Number Generator and seed value, etc) carried in the packet header is preferable.
%Moreover, thanks to heavy repair symbol, SRLC has efficient recovery performance at the cost of extra XOR operations as described in Section~\ref{sec:eval}.

%-------------------------------------------------------------------
\subsection{Parameter Settings for $D_{nonbin}$ and $D_{bin}$}
\label{subsec:add-structure}
Let us now determine the most appropriate values for the \{$D_{nonbin}$, $D_{bin}$\} tuple for a given $k$ and target average performance.
Of course:
\begin{itemize}
\item $D_{nonbin}$ should be as small as possible to reduce computation complexity, and
\item $D_{bin}$ should be as small as possible so that IT decoding performs well.
\end{itemize}
To find appropriate values, we did as described in Algorithm~\ref{algo:find-value}.
We set the target average overhead to $0.1\%$ (same value as in Fig.~\ref{fig:avg-ineff-nonbin_1_40_-rlc}) plus a security margin (set to $0.5$) so as to accommodate some fluctuations during the optimization process.
As a result, we obtain a table of \{$D_{nonbin}$, $D_{bin}$\} tuples, with an entry for each $k$ value.
Note that this table (not reproduced here) does not need to be sent to the receiver(s) as the \{$k$, $D_{nonbin}$, $D_{bin}$\} tuple is communicated at synchronization time to the receiver(s).
This provides additional flexibility since the the target code performance may be changed dynamically for the following transfers, at the discretion of the sender.

\begin{algorithm}
\caption{Finding the right \{$D_{nonbin}$, $D_{bin}$\} values.}
\label{algo:find-value}
\begin{algorithmic}[1]
	\STATE $target\_overhead \leftarrow 0.001$; /* for instance */
	\STATE $security\_margin \leftarrow 0.5 $;  /* for instance */
	\FOR{$k=2$ to $10000$}
		\STATE /* First of all, find $D_{nonbin}$ if $D_{bin}$ is set to $0.5$ */
		\STATE $D_{bin} \leftarrow 0.5$;
		\STATE $\textit{Get\_nb\_1\_coeffs}(D_{bin})$;
		\FOR{$nb\_nonbinary\_columns = 0$ to $k$}
			\STATE $Get\_average\_overhead ( k, $
			\STATE $\;\;\;\;\;\;\;\;\;\;\;\;\;\;\;\;\;\;\; nb\_nonbin\_column, \textit{nb\_1\_coeffs}) $;
			    \IF{($average\_overhead<target\_overhead*security\_margin$)}
				\STATE $Set\_D_{nonbin}(nb\_nonbinary\_columns)$;
				\STATE break;
			\ENDIF
		\ENDFOR

		\STATE /* Then find smallest $D_{bin}$ for the selected $D_{nonbin}$ */
		\WHILE{($true$)} 
			\STATE $\textit{nb\_1\_coeffs} --$;
			\STATE $Get\_average\_overhead ( k, $
			\STATE $\;\;\;\;\;\;\;\;\;\;\;\;\;\;\;\;\;\;\; nb\_nonbin\_column, \textit{nb\_1\_coeffs})$;
			\IF{($average\_overhead > target\_overhead$)}
				\STATE $Set\_D_{bin}(\textit{nb\_1\_coeffs}\;+\;1)$;
				\STATE break;
			\ENDIF	
			
		\ENDWHILE
	\STATE $store\_results(k, D_{nonbin}, D_{bin})$;
	\ENDFOR

\end{algorithmic}
\end{algorithm}

%-------------------------------------------------------------------
\subsection{Application to Convolutional Coding}
\label{subsec:appli-convol}

\begin{algorithm}
\caption{Building repair symbols in convolutional mode.}
\label{algo:convol}
\begin{algorithmic}[1]

\STATE $\textit{Alloc\_repair\_symbol\_buffer}(r)$;		/* reset to zero as well */
\STATE $\textit{Alloc\_heavy\_repair\_symbol\_buffer}(h)$;	/* reset to zero */
\WHILE{($true$)} 
	\STATE $Wait\_new\_src\_symbols()$;
	\STATE $Send\_new\_src\_symbols()$;
	\FORALL {(new source symbol $s$)}
		\STATE $h \leftarrow h \wedge s$;
	\ENDFOR
	\STATE $Set\_nb\_repair\_to\_send(total\_new\_src\_symbols,$
	\STATE $\;\;\;\;\;\;\;\;\;\;\;\;\;\;\;\;\;\;\;\;\;\;\;\;\;\;\;\;\;\;\;\; code\_rate)$;

	\WHILE {($nb\_repair\_to\_send > 0$)}
	\IF{($Decide\_to\_send\_heavy\_repair()$)}
		\STATE $Send\_repair\_symbol(h)$;
	\ELSE
		\STATE $Reset\_repair\_symbol\_memory(r)$;
		\STATE $Set\_new\_union\_of\_encoding\_windows(k)$;
		\FORALL {(src symbol $s$ in encoding window)}
			\IF {($(src\_symbol\_id \;\%\; D_{nonbin}) = 0$)}
				\STATE /* non-binary col., choose coeff randomly */
				\STATE $\textit{Set\_nonbin\_coefficient}()$;
				\STATE $r \leftarrow r \wedge (\textit{nonbin\_coefficient} * s)$;
			\ELSE
				\STATE /* binary column, choose 0 or 1 randomly */
				\STATE $\textit{Set\_binary\_coefficient}(D_{bin})$;
				\IF {($\textit{binary\_coefficient} =1 $)}
					\STATE $r \leftarrow r \wedge s$;
				\ENDIF
			\ENDIF
		\ENDFOR
		\STATE $r \leftarrow r \wedge h$;
		\STATE $Send\_repair\_symbol(r)$;
	\ENDIF
	\STATE $nb\_repair\_to\_send --$;
\ENDWHILE
%\IF {(The source symbol is the last)}
%	\STATE $end \leftarrow 1$
%\ENDIF	
\ENDWHILE

\end{algorithmic}
\end{algorithm}

Convolutional coding is appropriate to situations where a fully or partially reliable delivery of continuous data flows is needed, especially when these data flows feature real-time constraints, as in~\cite{tetrys}.
SRLC can then be used as a convolutional code, in a systematic way (i.e., source symbols are sent on the network), as described in Algorithm~\ref{algo:convol}.
The way the encoding window is managed (i.e., how to set the encoding window start and the number $k$ of source symbols  in the window) is a key aspect that depends on the protocol in use.

\begin{figure}[t]
    \begin{center}
    \includegraphics[width=200pt]{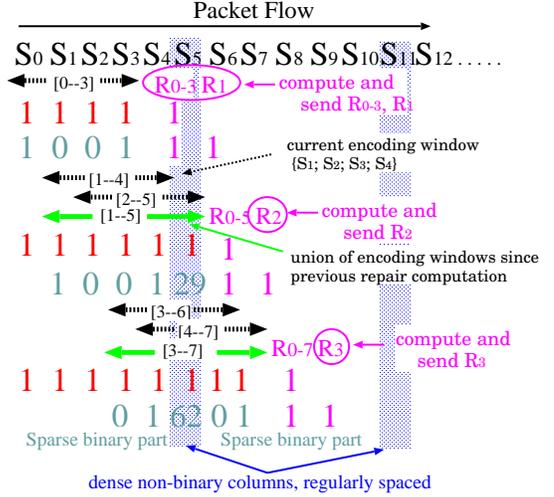}
    \end{center}
	\vspace{-3mm} %XXX
    \caption{SRLC example as a convolutional code, with fixed $k=4$ $CR=2/3$}
    \label{fig:srlc-convol-example}
\end{figure}

Fig.~\ref{fig:srlc-convol-example} illustrates the use of SRLC in the simple sliding window mode.
Here the encoding window has a fixed size, $k=4$, and slides in a regular way over the source symbol flow. 
The target code rate ($CR=2/3$ is such that one repair symbol is sent after two source symbols.
The only exception is at session start: the encoder waits for $k=4$ source symbols to be available, and then generates two repair symbols, including a heavy repair one, $R_{0-3}$ (i.e., the XOR sum from $S_0$ to $S_3$).
Then, after sending two more source symbols, $S_4$ and $S_5$, the SRLC encoder considers the union of the encoding windows since the previous repair computation (i.e., from $S_1$ to $S_5$) and generates a new repair symbol, $R_2$.
$R_2$ accumulates the current heavy repair symbol, $R_{0-5}$ (i.e., the XOR sum from $S_0$ to $S_5$) to the encoding vector:
	$$R_2 = S_1 + S_4 + 29*S_5 + R_{0-5}$$
Here also, the encoding vector is set according to the \{$k$, $D_{nonbin}$, $D_{bin}$\} tuple, using pre-calculated tables as described in Section~\ref{subsec:add-structure}, and a Pseudo-Random Number Generator (PRNG) that can be seeded by a specific value communicated to the receiver.
Note that the repair symbol identifier may be used as a seed.

From a signaling point of view, we can assume that the \{$k$, $D_{nonbin}$, $D_{bin}$\} tuple and all the algorithms are known by both ends. 
In that case, it is sufficient for the sender to let the receiver know the union of the encoding windows considered (e.g., from $S_1$ to $S_5$ in the case of $R_2$), the repair symbol identifier, along with the PRNG seed (if different from the repair symbol identifier).
This is all the SRLC decoder needs to know to generate the constraint equation associated to this repair symbol, even for large encoding window sizes.

%the tables and the knowledge of the algorithms of how to set constraint equations (e.g., $S_1 + S_4 + 29*S_5 $ in the case of $R_2$ in the example) to generate repair symbols are shared in advance by a sender and receiver, the sender only has to let the receiver know the scope of the union of the encoding window (e.g., from $S_1$ to $S_5$ in the case of $R_2$). Thus, it is sufficient for the sender to send only the parameters of the first $src\_symbol\_id$ in the encoding window and $k$ along with the repair symbol.
%If a seed value of Pseudo-Random Number Generator is used to set the constraint equations, the sender may add the seed value along with the repair packet, or the receiver may utilize a certain identifier (e.g., repair symbol identifier number).

In practice, the heavy repair symbols are transmitted periodically in order to remove the long term dependencies they create.
This is useful if past source symbols remain impossible to recover by a given receiver (who for instance joined the session late).
%Without the transmission of several values of the heavy repair symbol, the performance would seriously suffer as all the new repair symbols depend on them.

%==================================================================
\section{Performance Evaluation Results}
\label{sec:eval}
In this section we evaluate the SRLC erasure recovery performance both in block and convolutional modes.

%------------------------
\subsection{Experimental Setup}
%%% openfec, kodoをつかいました。
All the tests are carried out with the performance evaluation tools provided by our \verb+OpenFEC.org+ project~\cite{openfec_site} and a modified version of the \verb+Kodo+ library~\cite{kodo_site} for the codec implementation.
We use the pre-calculated values for \{$D_{nonbin}$, $D_{bin}$\} (see Algorithm~\ref{algo:find-value}) and we choose  $CR=2/3$ in all tests.
However SRLC is by nature rateless and the actual code rate is of little importance (i.e., the decoding overhead does not depend on the code rate).
Because we do not want to define any specific channel model (e.g., the two transition probabilities of a Gilbert model), in all tests we assume the source and repair symbols are transmitted in a fully random order, which means that only the packet loss rate is of importance.
Finally we assume that Gaussian Elimination decoding is used for maximum performance, rather than IT decoding (we do not consider decoding speeds in this work).

%------------------------
\subsection{Recovery Capabilities in Block Mode}

\begin{figure}[t]
    \begin{center}
    \includegraphics[width=220pt]{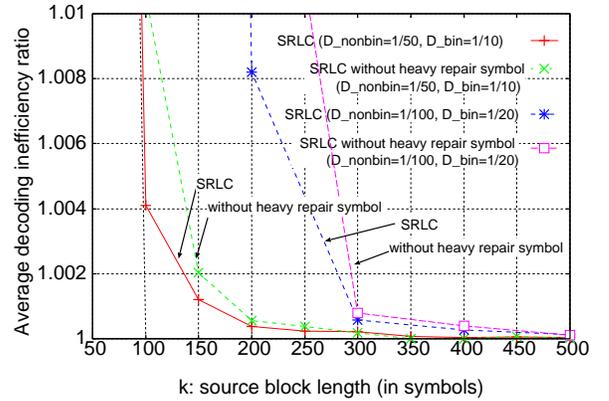}
    \end{center}
	\vspace{-3mm} %XXX
    \caption{On the benefits of heavy repair symbols: average recovery performance without and with (i.e., SRLC) this symbol.}
    \label{fig:avg-ineff-comp-heavyrepair}
\end{figure}

\begin{figure}[t!]
\centering
\subfigure[without heavy repair symbol]{
        \includegraphics[width=200pt]{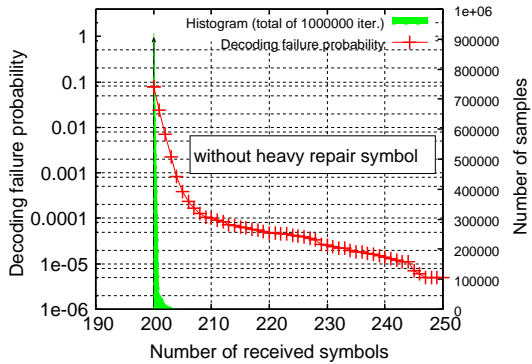}
        \label{fig:decfail-without-heavy}
}
\hfill
\subfigure[with heavy repair symbol]{
        \includegraphics[width=200pt]{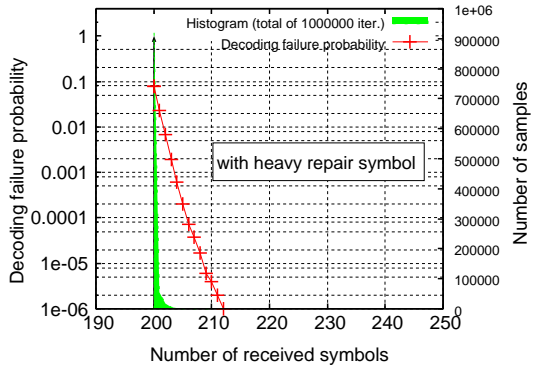}
        \label{fig:decfail-with-heavy}
}
\hfill
 \caption{Decoding failure probability when $k=200$ and ($D_{nonbin}$, $D_{bin}$) = ($1/50$, $1/10$)}
 \label{fig:srlc-decfail}
\end{figure}

Let us focus on the SRLC in block mode.
We measure both the average inefficiency ratio as a function of $k$ and the decoding failure probability as a function of the number of received symbols in addition to $k$
(in both cases decoding is said to fail as soon as at least one erased source symbol can not be recovered).
The first goal of tests is to demonstrate the efficiency of the use of a heavy repair symbol.
The second goal is to assess the performance of SRLC codes.
However, due to the space limitations, we only show the results when $k$ is small, from 50 to 500 symbols.

Fig.~\ref{fig:avg-ineff-comp-heavyrepair} compares the two options for $\{D_{nonbin}, D_{bin}\} = \{1/50, 1/10\}$ or $\{1/100, 1/20\}$.
We see the benefits of using the heavy repair symbol, especially when $k$ is small, on average.
Let us look at Fig.~\ref{fig:srlc-decfail}, when $k=200$ and $\{D_{nonbin}, D_{bin}\}=\{1/50, 1/10\}$.
In both cases, the decoding failure probability curves are similar when the number of received symbols is only slightly higher than $k$ (i.e., for low overheads).
However we clearly see a difference when the overhead is higher, meaning that there is a significant number of tests where decoding fails without any heavy repair symbol: 
$245$ extra symbols need to be received ($22.5$\% overhead) for the decoding failure probability to go below $10^{-5}$.
On the opposite, the full featured SRLC solution reaches a decoding failure probability lower than $10^{-5}$ with 209 symbols only (a $4.5$\% overhead).

Fig.~\ref{fig:srlc-decfail}-(b) also confirms the excellent recovery performance of SRLC codes, not only on average, but also when looking precisely at the decoding failure
probability.

%------------------------
\subsection{Recovery Capabilities in Convolutional Mode}
%Here, with realtime flows and convolutional blocks, we may admit some decoding failures
%from time to time. So the first two curves (src pkt loss ratio after FEC decoding) are perhaps more important than the decoding success proba curves (over the whole tot_src flow).

\begin{figure}[h]
\centering
\subfigure[Source Packet Loss Ratio (\%)]{
        \includegraphics[width=200pt]{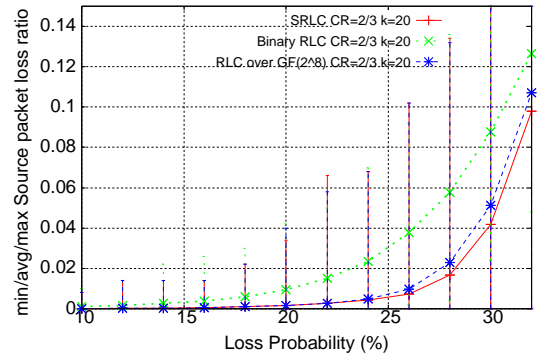}
        \label{fig:decfail-without-heavy}
}
\hfill
\subfigure[Decoding Failure Probability]{
        \includegraphics[width=200pt]{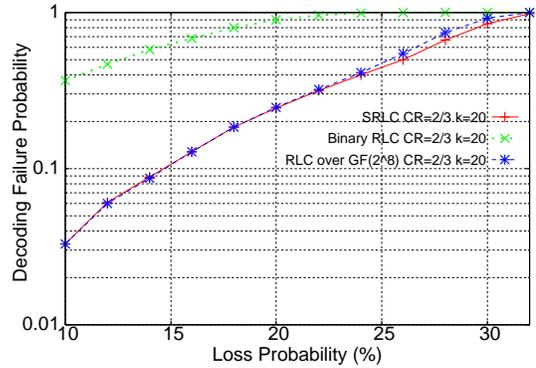}
        \label{fig:decfail-with-heavy}
}
\hfill
 \caption{Performances of SRLC, Binary RLC and RLC over $GF(2^8)$ in a sliding window (convolutional) mode when the total number of source symbols $tot\_src = 500$, encoding window size $k = 20$, $CR=2/3$.}
 \label{fig:srlc-srcloss-convol}
\end{figure}

\begin{figure}[h]
\centering
\subfigure[Source Packet Loss Ratio (\%)]{
        \includegraphics[width=200pt]{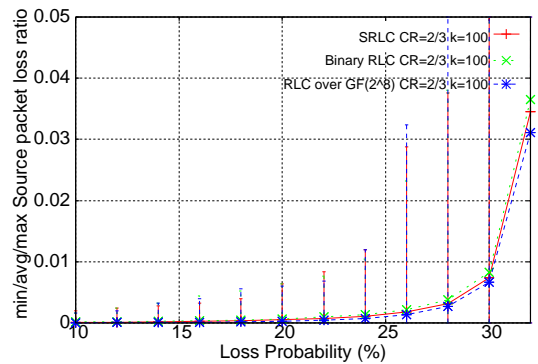}
        \label{fig:srcloss-k100}
}
\hfill
\subfigure[Decoding Failure Probability]{
        \includegraphics[width=200pt]{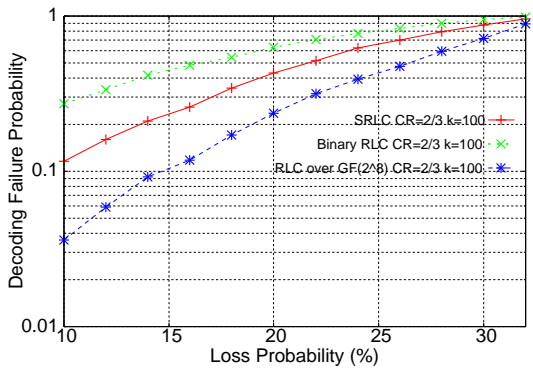}
        \label{fig:decfail-k100}
}
\hfill
 \caption{Performances of SRLC, Binary RLC and RLC over $GF(2^8)$ in a sliding window (convolutional) mode when the total number of source symbols $tot\_src = 2500$, encoding window size $k = 100$, $CR=2/3$.}
 \label{fig:srlc-decfail-convol}
\end{figure}

Let us now consider SRLC in convolutional mode.
Since we are focusing on real-time flows, like video/audio real-time streaming systems, a full reliability is not necessarily required (this is different from typical use in block mode, for file transfer applications).
Therefore we measure the SRLC average source packet loss ratio (once decoding is finished) as a function of packet loss probability, and compare it 
with those of binary RLC (i.e., $\{D_{nonbin}, D_{bin}\} = \{0, 1/2\}$) and of RLC over $GF(2^8)$ (all coefficients are randomly chosen in $GF(2^8)$).
Additionally, to make the comparison more visible, we measure the decoding failure probability of the three codes.

The performance results for the transmission of $500$ source symbols in total and a window of size $k=20$ symbols, are shown in Fig.~\ref{fig:srlc-srcloss-convol}.
We see that SRLC performs the best, even when compared to RLC over $GF(2^8)$, which is exceptional.
% because non-binary coefficients in SRLC are essential and the heavy repair symbol improves the performance at the cost of extra symbol operations.
%In that case, the decoding speed reduction caused by these extra operations can be ignored, because the small encoding window enables the decoder to solve smaller subsystems~\cite{rlc-raptor-comp}.

The performance results for a larger encoding window, of size $k=100$ symbols, are shown in Fig.~\ref{fig:srcloss-k100}.
We see that all the average loss ratios improve when compared to the $k=20$ case, because a larger encoding window size offers better protection.
Therefore there is no significant differences among the three codes especially on average.
However, when looking at the decoding failure probability in Fig.~\ref{fig:decfail-k100}, the SRLC performance pronouncedly becomes worse than that of RLC codes over $GF(2^8)$.
One reason is that SRLC uses the $\{D_{nonbin}, D_{bin}\}$ table optimized for the block mode case.
A new table should be calculated for the convolutional case.
%Instead, SRLC results in fewer symbol operations for decoding, which enables faster encoding/decoding.
%If a used application emphasizes the metric of decoding failure probability for a given loss probability, the parameter values can be changed.
%This implies that the parameter values in convolutional mode should be redefined according to the target recovery performances.

%==================================================================
\section{Conclusions and Future work}
\label{sec:concl}
This work introduces the SRLC codes, an end-to-end AL-FEC solution that is sufficiently flexible to be applied in block mode and convolutional mode.
In order to enable excellent erasure recovery performance as well as fast encoding and decoding speeds, these codes have been designed in a manner that favors a mostly
sparse and binary structure, with some well chosen non binary coefficients, plus a heavy binary row.
Additionally, the design is such that it facilitates an efficient signaling, the parameters exchanged to synchronize encoder and decoders being kept to a minimum.
These considerations make SRLC codes a very practical solution, no matter the block or encoding window size: small, medium or large.
Our evaluation of their erasure recovery performance confirms the benefits
%Additionally, these performances can easily be tuned to match the desired erasure recovery performance.

In future works we will analyze the encoding and decoding complexity (similarly the associated speeds) of SRLC codes.
We will also further optimize the value of the code internal parameters when used in convolutional mode, both in a fixed-size configuration and in elastic window configuration (e.g., as in~\cite{tetrys}).
%Finally, in IRTF Network Coding Research Group (NWCRG)~\cite{nwcrg}, we will standardize them to disseminate SRLC widely.
%Afterward, we plan to integrate SRLC into Asaeda CCN testbet~\cite{cutei} and evaluate it in terms of file/content acquisition time.

%==================================================================


\begin{thebibliography}{99}
\bibitem{rlnc}
T. Ho, R. Koetter,M. Medard, D. R. Karger and M. Effros,
``The Benefits of Coding over Routing in a Randomized Setting'',
IEEE International Symposium on Information Theory (ISIT'03), June 2003.

\bibitem{BATS1}
S. Yang and R. W. Yeung,
``Coding for a network coded fountain'',
IEEE International Symposium on Information Theory (ISIT'11), Aug. 2011.

\bibitem{BATS2}
T. Ng and S. Yang,
``Finite-length analysis of BATS codes'',
IEEE International Symposium on Network Coding (NetCod'13), June 2013.

\bibitem{flute}
T. Paila, R. Walsh, M. Luby, V. Roca, and R. Lehtonen,
``FLUTE - File Delivery over Unidirectional Transport'',
IETF RMT Working Group, Request for Comments, RFC 6726 ("Standards Track/Proposed Standard"), Nov. 2012.

\bibitem{ISDB-Tmm}
A. Yamada, H. Matsuoka, T. Ohya, R. Kitahara, J. Hagiwara, and T. Morizumi,
``Overview of ISDB-Tmm services and technologies'',
IEEE International Symposium on Broadband Multimedia Systems and Broadcasting (BMSM'11), June 2011.

%\bibitem{DVB-H}
%``ETSI TS 102 005,digital video broadcasting (dvb): Specification for the use of video and audio coding in dvb services delivered directly over ip protocols,''
%ETSI Tech. Spec, 2006.

\bibitem{MPE-FEC}
``Digital Video Broadcast (DVB); IP datacast: Content delivery protocols (cdp) implementation guidelines part 1: IP datacast over DVB'',
ETSI Technical Specifications, ETSI TR 102 591, 2007.

\bibitem{tetrys}
P.-U. Tournoux, E. Lochin, J. Lacan, A. Bouabdallah, and V. Roca,
``On the fly erasure coding for time-constrained applications'',
IEEE Transactions on Multimedia, vol. 13, no. 4, pp. 797-812, Aug. 2011.

\bibitem{rlc-raptor-comp}
S. Nazir, D. Vunkobratovi\' and V. Stankovi\'c,
``Peformance evaluation of Raptor and Random Linear Codes for H.264/AVC video transmission over DVB-H networks'',
IEEE International Conference on Acoustics, Speech and Signal Processing (ICASSP'11), May 2011.

\bibitem{linear-net-coding}
R.S.-Y.Li, R. W. Yeung, and N. Cai,
``Linear Network Coding'',
IEEE Transactions on Information Theory, vol. 49, no. 2, pp. 371-381, Feb. 2003.

\bibitem{digital-fountain}
J. Byers, M. Luby and M. Mitzenmacher,
``A digital fountain approach to asynchronous reliable multicast'',
IEEE Journal on Selected Areas in Communications vol. 20, No. 8, pp. 1528-1540, 2002.

\bibitem{vincent-wimob13}
V. Roca, M. Cunche, C. Thienot, J. Detchart and J. Lacan,
``RS+LDPC-Staircase codes for the erasure channel: Standards, Usage and Performance'',
IEEE International Conference on Mobile Computing, Networking and Communications (WiMob'13), Nov. 2013.

\bibitem{raptor}
A. Shokrollahi,
``Raptor codes'',
IEEE Transactions on Information Theory, vol. 52, no. 6, pp. 2251-2567, 2006.

\bibitem{gamma1}
K. Mahdaviani, M. Ardakani, H. Bagheri, and C. Tellambura,
``Gamma codes: a low-overhead linear-complexity network coding solution'',
International Symposium on Network Coding (NetCod’12), June 2012.

\bibitem{gamma2}
K. Mahdaviani, R. Yazdani, and M. Ardakani, 
``Overhead-optimized gamma network codes'',
International Symposium on Network Coding (NetCod’13), June 2013.

\bibitem{SGE1}
B. A. LaMacchia and A. M. Odlyzko,
``Solving large sparse linear systems over finite fields'',
Advances in Cryptology (Crypto'90), LNCS 5357, Springer-Verlag, 1991.

\bibitem{SGE2}
C. Pomerance and J. W. Smith,
``Reduction of huge, sparse matrices over finite fields via created catastrophes'',
Experimental Mathematics, vol. 1, no. 2, 1992.

\bibitem{openfec_site}
``OpenFEC.org: because open, free AL-FEC codes and codecs matter'',
http://openfec.org.

\bibitem{nsrlc.ietf88}
K. Matsuzono and V. Roca,
```Not so random RLC AL-FEC codes'',
http://hal.inria.fr/hal-00879834/en/,
NWCRG (NetWork Coding Research Group) meeting, IETF 88, Nov. 2013.


\bibitem{itdecoding}
V. Zyablov and M. Pinsker,
``Decoding complexity of low-density codes for transmission in a channel with erasures'',
Translated from Problemy Predachi Informatsii, 10(1), 1974.

\bibitem{ldpc-rfc}
V. Roca, C. Neumann and D. Furodet,
``Low Density Parity Check (LDPC) Staircase and Triangle Forward Error Correction (FEC) Schemes'',
IETF RMT Working Group, RFC 5170 ("Standards Track/Proposed Standard"), June 2008.

\bibitem{kodo_site}
``Kodo network coding library'',
https://kodo.readthedocs.org/en/latest/.


\bibitem{fecframework}
M. Watson, A. Begen and V. Roca,
``Forward Error Correction (FEC) Framework'',
IETF FECFRAME Working Group, RFC 6363 ("Standards Track/Proposed Standard"), ISSN 2070-1721, Oct. 2011.

\bibitem{nwcrg}
IRTF Network Coding Research Group (NWCRG),
http://irtf.org/nwcrg.

\bibitem{3GPP-MBMS}
3GPP Technical Specification Group Services and System Aspects,
``Multimedia Broadcast/Multicast Service (MBMS); Protocol and codecs (Release 6)'',
3GPP TS 26.346 v6.4.0, Mars 2006.

\bibitem{IRA}
H. Jin, D. Khandekar, and R. J. McEliece,
``Irregular repeat-accumulate codes'',
Second International Symposium on Turbo Codes and Related Topics, Brest, France, Sept. 2000.


%\bibitem{tcpfriendly-rap}
% R. Rejaie, M. Handley, and D. Estrin,
% ``RAP:An End-to-end Rate-based Congestion Control Mechanism for Realtime Streams in the Internet,'' Proc. of IEEE INFOCOM 1999, March 1999

%\bibitem{dvtsexample}
%  N. Nakashima, K. Okamura, JS. Hahm, YW. Kim, H. Mizushima, H. Tatsumi, BI. Moon, HS. Han, YJ. Park, JH. Lee, SK. Youm, CH. Kang, and S. Shimizu, ``Telemedicine with digital video transport system in Asia-Pacific area,''
% Proc. of the 19th International Conference on Advanced Information Networking and Applications, March 2005.

%\bibitem{rtp}
%  H. Schulzrinne, S. Casner, R. Frederick, and V. Jacobson,
%  ``RTP: A Transport Protocol for Real-Time Applications,''
%  IETF RFC 3550, July 2003.

%\bibitem{fec1}
%  J.C. Bolot and A. V. Garcia,
%  ``Control Mechanisms for Packet Audio in the Internet,''
%  Proc. of IEEE INFOCOM, March 1996.

%\bibitem{fec2}
%  C. Perkins and O. Hodson,
%  ``Options for Repair of Streaming Media,''
%  IETF RFC 2354, June 1998.

%\bibitem{rfc3453}
%  M. Luby, L. Vicisano, J. Gemmell, L. Rizzo, M. Handley, and J. Crowcroft,
%  ``The Use of Forward Error Correction (FEC) in Reliable Multicast,''
%  IETF RFC 3453, December 2002.

%\bibitem{rs-code}
%  L. Rizzo,
%  ``Effective erasure codes for reliable computer communication protocols,''
%  ACM Comput. Commun, vol.27, no.2, pp.24-36, April 1997.

%\bibitem{rse-rfc}
%  J. Lacan, V. Roca, J. Peltotalo, and S. Peltotalo,
%  ``Reed-Solomon Forward Error Correction (FEC) Schemes''
%  IETF RFC 5510, April 2009.

%\bibitem{ldpc1960}
%  D. MacKay,
%  ``Information Theory, Inference and Learning Algorithms,''
%  Cambride University Press, ISBN : 0521642981, 2003.


%\bibitem{ldpc-hybrid-inria}
%  M. Cunche and V. Roca,
%  ``Improving the ecoding of LDPC codes for the packet erasure channel with a hybrid Zyablov iterative decoding/Faussian elimination scheme,''
%  Research Report 6473, INRIA, March 2008.

%\bibitem{mathieuspsc08}
%  M. Cunche and V. Roca,
%  ``Optimizing the Error Recovery Capabilities of LDPC-staircase Codes Featuring a Gaussian Elimination Decoding Scheme,''
%  Proc. of IEEE International Workshop on Signal Processing for Space Communications (SPSC'2008).

%\bibitem{dvts}
%  A. Ogawa, K. Kobayashi, K. Sugiura, O. Nakamura, and J. Murai,
%  ``Design and Implementation of DV based video over RTP,''
%  Proc. of Packet Video Workshop 2000, May 2000.

%\bibitem{dvtsrfc}
%  K. Kobayashi, A. Ogawa, S. Casner, and C. Bormann.
%  ``RTP Payload Format for DV (IEC 61834) Video,''
%  IETF RFC 3189, January 2002.

%\bibitem{rsvp}
%  R. Braden, Ed., L. Zhang, S. Berson, S. Herzog, and S. Jamin,
%  ``Resource ReSerVation Protocol (RSVP) -- Version 1 Functional Specification,''
%  RFC 2205, September 1997.

%\bibitem{cbq}
%  S. Floyd and V. Jacobson,
%  ``Link-sharing and Resource Management Models for Packet Networks,''
%  IEEE/ACM Transactions on Networking, Vol.3, No.4, pp.365-386, August 1995.

%\bibitem{ecn}
% S. Floyd,
% `` TCP and explicit congestion notification,''
% ACM Comput. Commun, Vol.24, No.5, pp.10-23, Oct 1994.

%\bibitem{tcpfriendly1}
%  S. Floyd and K. Fall,
%  ``Promoting the Use of End-to-End Congestion Control in the Internet,''
%  IEEE/ACM Transactions on Networking, Vol.7, No.4, pp.458-472, August 1999.

%\bibitem{equationuni}
% S. Floyd, M. Handley, J. Padhye, and J. Widmer,
% ``Equation-Based Congestion Control for Unicast Applications,''
% Proc. of ACM SIGCOMM '00, August 2000.

%\bibitem{adaptivempeg}
% P. Papadimitriou and V. Tsaoussidis,
% ``A Rate Control Scheme for Adaptive Video Streaming Over the Internet,''
% Proc. of IEEE ICC '07, June 2007.

%\bibitem{tcpfriendly2}
%  M. Handley, S. Floyd, J. Padhye, and J. Widmer,
%  ``TCP Friendly Rate Control (TFRC): Protocol Specification,''
%  IETF RFC 3448, January 2003.

%\bibitem{sigcomm97}
%  V. Paxson,
%  ``End-to-End Internet Packet Dynamics,''
%  Proc. of ACM SIGCOMM '97, pp.139-152, 1997, Cannes, France.

%\bibitem{nossdav95}
%  J.C. Bolot, H. Cr\'epin, and A. Vega-Garcia,
%  ``Analysis of Audio Packet Loss over Packet-Switched Networks,''
%  Proc. of ACM NOSSDAV '95,
%  April 1995, New Hampshire, USA.

%\bibitem{nossdav05}
%  H. Wu, M. Claypool, and R. Kinicki,
%  ``Adjusting Forward Error Correction with Quality Scaling for Streaming MPEG,''
%  Proc. of ACM NOSSDAV '05, June 2005, Washington, USA.

%\bibitem{ton05}
%  I. F. Akyildiz, \"O. B. Akan, and G. Morabito,
%  ``A rate control scheme for adaptive real-time applications in IP networks with lossy links and long round trip times,''
%  IEEE/ACM Transactions on Networking, Vol.13, Issue 3, pp.554-567, June 2005.

%\bibitem{dccp}
%  E. Kohler, M. Handley, and S. Floyd, Designing ^[$B!H^[(BDCCP: Congestion
%  control without reliability^[$B!I^[(B, in Proc. ACM SIGCOMM 2006,
%  September 2006.

%\bibitem{limitp2p}
% PeerApp White Paper.
% ``Comparing P2P solutions,''
% http://www.peerapp.com/Data/Files/ComparingP2PSolutions.pdf, March 2007.

%\bibitem{fasttcp}
% D.X. Wei, C. Jin, S.H. Low, S. Hegde,
% ``FAST TCP: Motivation, Architecture, Algorithms, Performance,''
% IEEE/ACM Transactions on Networking, Vol.14, NO.6, December 2006.

%\bibitem{cc-paradigm}
% A. Legout, and E.W. Biersack,
% ``Revisiting the fair queuing paradigm for end-to-end congestion control,''
% IEEE NetWORK, pp.38-46, oct 2002.

%\bibitem{myglobecom}
% K. Matsuzono, K. Sugiura, and H. Asaeda,
% ``Adaptive Rate Control with Dynamic FEC for Real-Time DV streaming,''
% Proc. of IEEE Globecom'08, December 2008.

%bitem{starbed}
%``StarBED Project,'' http://www.starbed.org/

%\bibitem{aimd}
% D. Chiu and R.Jain,
% ``Analysis of the increase and decrease algorithm for congestion avoidance in computer networks,''
% Comput. Netw. ISDN Syst. J., vol.17, NO.1, pp.1-14, June 1989.

%\bibitem{ns2}
%``The Network Simulator -- ns-2,'' http://www.isi.edu/nsnam/ns

%\bibitem{dummynet}
%  L. Rizzo, dummynet, http://info.iet.unipi.it/\~{}luigi/ip\_dummynet/

%\bibitem{ratedv}
%  Xin WANG and Henning SCHULZRINNE,
%  ``Comparison of Adaptive Internet Multimedia Applications,''
%  IEICE Tran. COMMUN., Vol.E82-B, NO.6, June 1999.

%\bibitem{planetlab}
%  B. Chun, D. Culler, T. Roscoe, A. Bavier, L. Peterson, M. Wawrzoniak, and M. Bowman,
%  ``PlanetLab: An Overlay Testbed for Broad-Coverage Services,''
%  ACM SIGCOMM Computer Communication Review, Vol.33, No.3, July 2003.

%\bibitem{aintec}
%  K. Matsuzono, H. Asaeda, K.Sugiura, O. Nakamura, and J. Murai,
%  ``Analysis of FEC function for Real-Time DV streaming,''
%  Asian Internet Engineering Conference '07, LNCS 4866, pp.114-122,
%   Novemver 2007.

%\bibitem{dvtsurl}
%``Digital Video Transport System -- DVTS,'' http://www.sfc.wide.ad.jp/DVTS/

%\bibitem{globecom2004}
% L. Roychoudhuri and E.S. Al-Shaer,
% ``Adaptive Rate Control for Real-time Packet Audio Based on Loss Prediction,''
% Proc. of IEEE Globecom '04

%\bibitem{reno}
%  M. Allman, V. Paxson, and W. Stevens,
%  ``TCP Congestion Control,'' IETF RFC 2581, April 1999.

%\bibitem{nossdav-dvts}
%  C. Bao, X. Li, and J. Jiang,
%  ``Scalable Application-Specific Measurement Framework for High Performance Network Video,''
%  Proc. of ACM NOSSDAV '07, Urbana, Illinois USA.

%\bibitem{adaptive-fec-infocom99}
% J.C. Bolot, S. Fosse-Parisis, and D. Towsley,
% ``Adaptive FEC-based error control for Internet telephony,''
% Proc. of IEEE INFOCOM '99, New York, USA, March 1999.

%\bibitem{pk-schedule-fec}
% C. Neumann, V. Roca, A. Francillon, and D. Furodet,
% ``Impacts of Packet Scheduling and Packet Loss Distribution on FEC Performances: Observations and Recommendations,''
% Proc. of CoNEXT'05, Toulouse, France, October 2005.

%\bibitem{switch-fec}
% C. Lamoriniere, A. Nafaa, and L. Murphy
% ``Dynamic Switching Between Adaptive FEC Protocols for Reliable Multi-Source Streaming,''
% Proc. of IEEE Globecom'09 Hawai USA, November 2009.

%\bibitem{dvts_site}
% "DVTS: Digital Video Transport System, or DV Stream on IEEE1394 Encapsulated into IP",
% http://www.sfc.wide.ad.jp/DVTS/.


\end{thebibliography}
\end{document}